\documentclass[sigconf]{acmart}
\renewcommand\footnotetextcopyrightpermission[1]{} 
\usepackage{graphicx}
\usepackage[lofdepth,lotdepth]{subfig}
\usepackage{amssymb}
\usepackage{fancyhdr}
\AtBeginDocument{%
  \providecommand\BibTeX{{%
    \normalfont B\kern-0.5em{\scshape i\kern-0.25em b}\kern-0.8em\TeX}}}

\begin{document}

\title{Effectively Using Long and Short Sessions for Multi-Session-based Recommendations }

\author{Zihan Wang}
\affiliation{%
  \institution{Northeastern University }
  \country{China}}
\email{2101816@stu.neu.edu.cn}
\author{Gang Wu}
\affiliation{%
  \institution{Northeastern University }
  \country{China}}
\email{wugang@mail.neu.edu.cn}
\author{Yan Wang}
\affiliation{%
  \institution{Northeastern University }
  \country{China}}
\email{orangewy97@gmail.com}

\begin{abstract}
  It is not accurate to make recommendations only based one single current session. Therefore, multi-session-based recommendation(MSBR) is a solution for the problem. Compared with the previous MSBR models, we have made three improvements in this paper. First, the previous work choose to use all the history sessions of the user and/or of his similar users. When the user's current interest changes greatly from the past, most of these sessions can only have negative impacts. Therefore, we select a large number of randomly chosen sessions from the dataset as candidate sessions to avoid over depending on history data. Then we only choose to use the most similar sessions to get the most useful information while reduce the noise caused by dissimilar sessions. Second, in real-world datasets, short sessions account for a large proportion. The RNN often used in previous work is not suitable to process short sessions, because RNN only focuses on the sequential relationship, which we find is not the only relationship between items in short sessions. So, we designed a more suitable method named GAFE based on attention to process short sessions. Third, Although there are few long sessions, they can not be ignored. Not like previous models, which simply process long sessions in the same way as short sessions, we propose LSIS, which can split the interest of long sessions, to make better use of long sessions. Finally, to help recommendations, we also have considered users' long-term interests captured by a multi-layer GRU. Considering the four points above, we built the model ENIREC. Experiments on two real-world datasets show that the comprehensive performance of ENIREC is better than other existing models.
\end{abstract}


\keywords{multi-session-based recommendation, attention, long short term interest}


\maketitle

\section{Introduction}
Generally speaking, users' interests are by no means invariable. Important factors such as age, work, scientific and technological development have great influences on users' interests. And capturing the changes of users' interests is a key point. The traditional methods are not good enough at this point. Session-based recommendation(SBR) come into being to solve this problem\cite{3}. In SBR, a user's behaviors are divided into multiple sessions, and behaviors in each session occurs in a short period of time. In this way, we turn a user's interest into many sessions' interests. So, each session of a user can present his interest during that time. This makes changes of user interests easier to find. Sessions are all serialized data and there are many ways to process serialized data. These methods Can achieve good results when dealing with long sessions. But the results of these methods in the recommendation system are always unsatisfactory. This is because most of the sessions of the recommendation datasets are short sessions (generally, sessions with a length of less than or equal to four are defined as short sessions). The proportion of short sessions in the commonly used datasets $Delicious$ and $Reddit$ of the recommendation system is 64.03\% and 96.95\% respectively. Because there are only a few items in a short session, it is difficult to give accurate prediction only base on a single current session.

To solve this problem, multi-session-based recommendation(MSBR) is a good method. The core idea is that when we can't get a long enough session, we can find multiple similar sessions to help recommendation. Previous methods have considered finding similar sessions from session sets such as users' history sessions and/or similar users' history sessions, but their approach has two problems. (\textbf{PROBLEM1}) They did not reduce the huge noise caused by multi session, and these data are too dependent on history, which sometimes interferes with our judgment. We need a new strategy for selecting sessions, which should avoid completely relying on history and only choose to use sessions that can really have a positive impact.

As we mentioned before, traditional methods can not achieve good results in short sessions. (\textbf{PROBLEM2}) So we need solution for the problem of short session processing. At the same time, although the number of long sessions is small, it can not be ignored, and long sessions and short sessions obviously have different characteristics. (\textbf{PROBLEM3}) Therefore, we also need to design a set of processing methods for long sessions. This processing method is better when it can associate with the processing method of short sessions above.

By solving the above three problems, we can improve existing MSBR. For PROBLEM1, in order to avoid over reliance on historical data and increase the robustness of the model, we not only use the sessions of users and similar users, but also choose to obtain a large number of sampled sessions from the datasets. Then we calculate the similarity with the current session for each selected session. According to the similarity, we can find the most similar sessions, which can also provide the most positive help to the recommendation. For PROBLEM2, We found that the attention mechanism\cite{1} can extract context relations, which is more appropriate for short sessions\cite{20}. So we designed GRU-ATTENTION Feature Extractor(GAFE), which takes into account the advantages of RNN and ATTENTION, and is more suitable for short sessions. And For PROBLEM3 , we designed Long Session Interest-Spliter(LSIS), which can divides a long session into multiple short sessions by sliding window. In this way, the interest of long sessions can be changed into short sessions with more targeted interest. Then we can use GAFE to process these short sessions, too. 

Finally, in order to improve the recommendation performance, we also consider that users' long-term interests are an important information. Specially, in SBR, long-term interests are more likely to contain more information in changes of interests\cite{19,28}. For example, if the items of a user's past session are high school books, and then another session later contains some college books. An implicit message is that the user was a high school student and is now a university student. Therefore, we also design a multi-layer GRU to extract users' long-term interest.

By solving the above problems, we created our model Enhance-Next-Item Recommender(ENIREC), it is divided into four modules: (1)Sim Sessions' Interest Module for make good use of similar sessions' interest, (2) User Current Interest Module to learn what the user want now, (3)Long-Term Interest Module to extract his long-term interest, (4)Prediction Module to make prediction base on the three parts of information above. We summarize the contribution of our work into the following three points:

\begin{itemize}

\item We find that the existing MSBR model has some problems in selecting sessions. In order to solve these problems. we improved the selection of sessions. It greatly retains the advantages of the MSBR and avoids the disadvantages.

\item There is a flaw in previous session processing methods. We designed GAFE, a more suitable method to deal with short sessions. And a scheme of GAFE+LSIS is designed to deal with long sessions. So, We can achieve better results in the processing of sessions.

\item We also designed a module to capture the change of users' long-term interests. Considering current session interest, sim sessions interest and long-term interest, we designed ENIREC for recommendations. Extensive experiments on two real datasets with a high propotion of short sessions proved that ENIREC is better than state-of-the-art models in this kind of datasets.

\end{itemize}

\begin{figure}[ht]
 \centering
 \includegraphics[height=0.22\textheight]{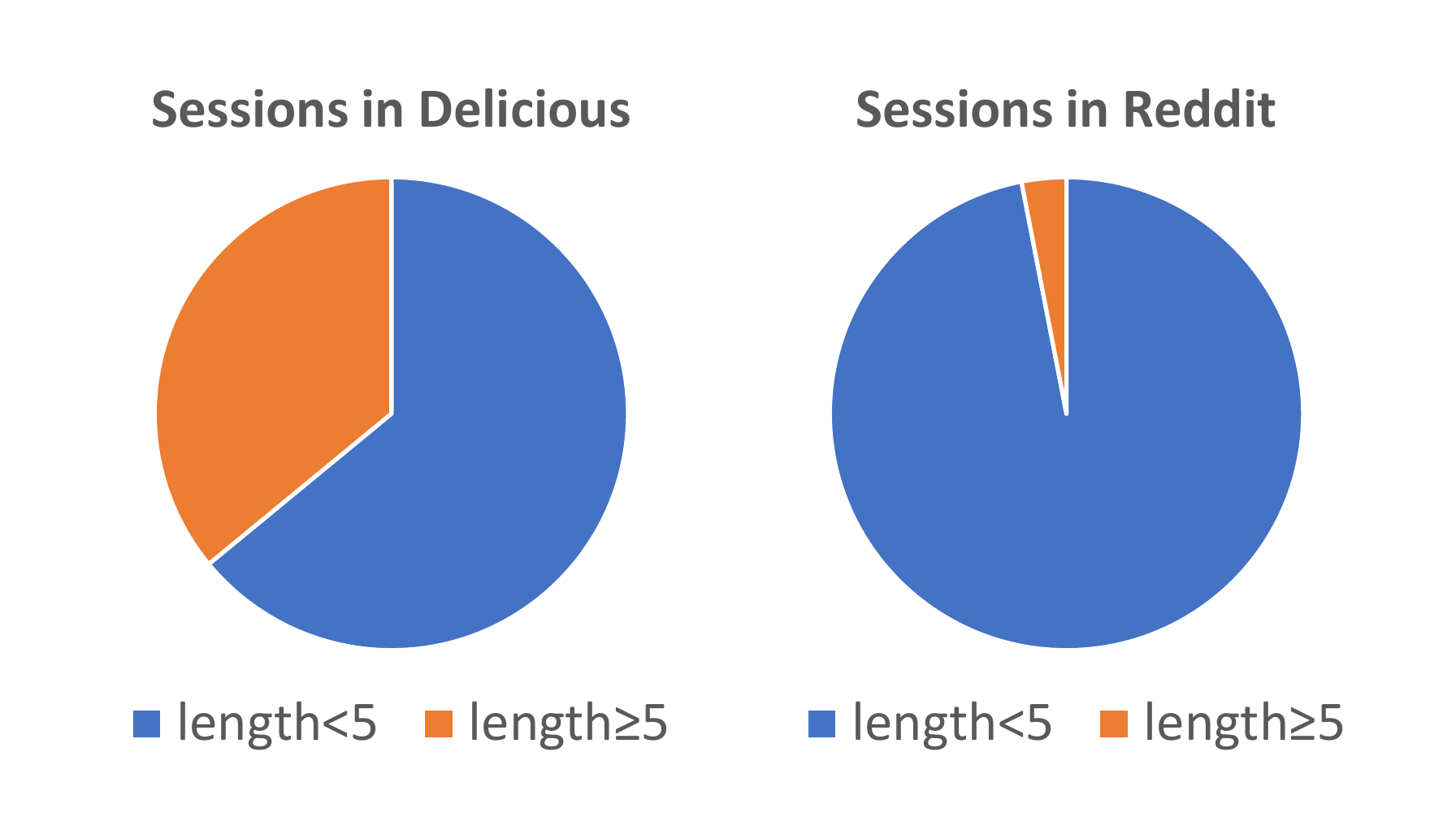}
 \caption{Propotion of short sessions}
 \label{fig:DEANDRE}
\end{figure}

\section{RELATED WORK}

\subsection{Multi-Session-Based Recommendation}

The obvious problem of short sessions is that the amount of available information is small. The most basic principle we follow in recommendation is that more information brings more accuracy. In the previous sequence processing, a long sequence contains enough items. When we extract the sequence relationship, they bring a lot of information, which will greatly improve the accuracy of recommendation. However, in a short sequence whose sequence length does not exceed five, it is obvious that there is not enough information. Since we can't increase the length of one session, we can only consider to collect more information from other sessions. In a word, more sessions, more information. This allows us to focus our work from Single-Session Based Recommendation(SSBR)\cite{26} to Multi-Session-Based Recommendation(MSBR). 
In the real-world datasets, we can often find information from hundreds of other sessions rather than only one current session, which means that we have also expanded the amount of our data by so many times. A paper propose to find similar sessions from the whole datasets by clustering\cite{26}. At the begin, scholars find similar sessions from the user's history sessions\cite{16}, but the sessions found in this way are still not enough. Similar users are also the information often considered in the recommendation system\cite{23}. Therefore, they begin to consider finding similar sessions from similar users\cite{24}. The latter two methods are more valuable because they are more targeted, so they can find more similar sessions.

But in fact, most candidate sessions are still irrelevant\cite{2}. If all these sessions are used blindly, it will bring great noise. Therefore, we need to find the most relevant sessions in these sessions and make selective use of them\cite{27}. The two type of sessions are actually selected according to a user's history behaviors( Similar users are also found according to the user's history) . Once the user's interest changes, these sessions will cause huge noise and affect the recommendation. 

Therefore, we need to be very careful in the selection of similar session to reduce the possible impact of noise in MSBR.

\subsection{RNN in Recommendation}

We also need to consider how to use these sessions. The simplest and most commonly used way is to treat the sessions as sequence data and extract information by some sequence processing methods. This has indeed proved useful in some experiments.

Recurrent Neural Network(RNN) is a complex network structure. RNN is a model proposed to capture the sequence relationship. RNN will not only consider the information of nearest layer of the sequence, but also consider the output of the previous layer. In this way, RNN can accurately model the precedence relationship. 

Therefore, RNN such as GRU is unique in processing sequence data. The essence of SBR is to process session sequences. So people think that the application of RNN to SBR should achieve good results. 

GRU4Rec\cite{5} proposed some time ago seems to have achieved good results, but it is still not good enough. The success of GRU4Rec shows that RNN can still achieve good results for sequence data in SBR, but RNN can not be the final solution of SBR. Because RNN strictly depends the order, while there is not only an order relationship implied in the session. For example, in the classic case of Wal-Mart ,the beer and diapers, a man often buys beer and diapers at the same time, but the two things are not in order. It is possible for him to buy either one first. This suggests that we need some new methods. Natural Language Processing(NLP) and Sequential Recommendation have great similarities. Some methods for NLP are also used in recommendation. For example, recent years, Transformer from NLP\cite{7} has also been used in the field of recommendation and achieved good results. The success of Bert4Rec\cite{18} proves that transformer is also an effective scheme\cite{22}. The Graph Neural Network(GNN) method in Image Processing is also used for recommendation\cite{6,10}, it connects a single sequence with other sequences as a graph. The available data has changed from sequence data to richer graph data, which enriches the acquisition of information. Some people even mixed many methods to build models\cite{9,17,30}. However, the sessions in SBR may not be able to give full play to the advantages of these methods. When the training data are mostly very short sessions, the effect of learning in these models is not satisfied. The inspiration from this result is that we should adopt some new methods to deal with short sequences. Moreover, completely changing session processing into another form of problem will also bring serious damage to the session information structure\cite{14}.

The conclusion in that all these methods for recommendation have one obvious thing in common, that is, they are not short sessions solutions. When dealing with datasets with a large number of short sessions, it is stretched.

\section{Problem Formulation}

In the Sequential recommendation problem, there are three sets of data needed to define the problem: users, items, and sessions. We use U, I and S to represent them respectively. $ U=\left\{ u_1,u_2...,u_{\lvert U \rvert}\right\} $ , $ I=\left\{ i_1,...,i_{\lvert I \rvert}\right\} $ , the U set contains all users, and all items are in the I set. The relationship between sessions and items is that one session contains multiple items. The content of a single session is like $ s_t=\left\{i_1^{s_t},...,i_{\lvert s_t \rvert}^{s_t}\right\} $ and Sessions set $ S =\left\{ s_1,s_2...,s_{\lvert S \rvert}\right\} $. One user have some sessions $ U_j=\left\{s_1^{U_j},...,s_{\lvert {U_j} \rvert}^{U_j}\right\} $ . Tensors used to represent features in this paper are all represented by $e = [a_1,...,a_n]$ with dimension $n$ .
\begin{itemize}

\item {\verb| Pre-Trained GRU |}: The purpose of the Pre-Trained Module is to obtain a generally stable Recurrent Neural Network with gated current units (GRU). We will use the sessions in the current training set to pre-train the GRU .We will use this GRU in the following modules, so as to make the prediction of the model more accurate in the initial step, and the GRU will also be automatically fine-tuned\cite{29} in the subsequent model training.We choose sessions in the train set $ S_{train} =\left\{  s_1^{train},...,s_{\lvert S_{train} \rvert}^{train}\right\} $ , for sessions in $ S_{train} $ like $s_1^{train}=\left\{{i_1^{s_1^{train}},...,i_{\lvert s_1^{train} \rvert}^{s_1^{train}}}\right\}$ , we use $\left\{i_1^{s_1^{train}},...,i_{\lvert s_1^{train} \rvert-1}^{s_1^{train}}\right\}$ to predict the last item $ i_{\lvert s_1^{train} \rvert}^{s_1^{train}} $ for a pre-trained GRU.

\item {\verb| Long-term Interest Module |}: In order to capture users' long-term interests, we need to pay attention to capturing users' interests in each session in the past. By making good use of the interests of each session, we can get the long-term interests of users. The method we choose is to regard the interests of these sessions as a change sequence of interests in chronological order $\left\{e_1^{history},...,e_{\lvert S_{history} \rvert}^{history}\right\}$. Extracting the interests of this change sequence by sequence method is the long-term interests of users. To utilize the interest sequence to represent users' long-term interest. We build a multi-layer GRU based on it to extract users' long-term interests. The output of this Module is user's long-term interest $E_{long-term-interest}$ .

\item {\verb| Current Interest Module |}: The user's current session has the most important information. We must extract the current interest of a user from his current session $s_{current} = \left\{ i_1^{current},...,i_{\lvert s_{current} \rvert}^{current}\right\}$ for recommendation and other aspects of the model. For example, in the part of calculating sessions' similarities, the reference hormone for similarity are the users' current interests. Current Interest Module want to get user's current interest. We can simply use GAFE to process user's current session. The output of this GAFE is user's current interest $E_{current-interest}$ .

\item {\verb| Sim Session's Interest Module |}: The main function of the Sim Session's Interest Module is to extracted information from chosen sessions. In order to provide it to subsequent Prediction Module. These sessions comes from three parts : randomly sampled similar sessions from the dataset  $S_{sample}$ , sessions from user's history $S_{history}$ and his similar user's history sessions $S_{simuser}$ . For long sessions in three sets, we use Long Session Interest Spliter(LSIS) and GRU-ATTENTION Feature Extractor(GAFE) to expand the data and increase the amount of information we can use. Use GAFE to process short sessions directly. Eventually, we will get the feature representation of these sessions  $ E_{sim}=\left\{ e_1^{E_{sim}},...,e_{\lvert E_{sim} \rvert}^{E_{sim}}\right\} $. Then inner -product these vectors with the user's current interest $E_{current}$ to get similarities $ Sim_{li}=\left\{Sim_1,..., Sim_{\lvert Sim_{li}\rvert}\right\}$. Multiply the tensors in $ E_{sim} $ by their corresponding similarity in $ Sim_{li} $one by one . Finally, Add up these output results. We can have the output of this module $E_{sim-interest}$

\item {\verb| Predict Module |}: The task of the Predict Module is to integrate the outputs of Long-term Interest Module,  Current Interest Module, Sim Session's Interest Module three modules . Through a Fully Connected Layer, we can have the next-item interest $ E_{next-item}$ , and then inner-product with the embedding of all items $ E_{items}=\left\{ e_{i_1},...,e_{i_{\lvert I \rvert}}\right\} $to get the score. Finally, select the top-k items with the highest score for the users and recommend it to them.

\end{itemize}

\section{ENHANCE NEXT-ITEM RECOMMENDATION}
We will introduce our model ENHANCE NEXT-ITEM RECOMMENDATION(ENIREC) in detail in this section. At some important parts, we use formulas to express the mathematical process. 
First of all, we need to make it clear that our model is divided into four modules, of which three modules are used to prepare information for prediction, and the last Prediction Module will predict next-item for users based on the information provided by these three modules. Three modules provide three aspects of information. Current Interest Module provide users' current interests. Long-term Interest Module provide users' long-term interests. We all know the importance of these two interests. They are the two data most relevant to users. Therefore, we must design appropriate modules for these two interests. 

The last module, Sim session's Interest Module is the core of our whole model, and the information it extracts is also very critical. Similar to Few-Shot Learning, Short-Session Based Recommendation is also base on a small amount of data. The solution for this kind of problem is different from the prediction with a lot of information. The solution of this kind of problem has changed from how to accurately find an item to minimizing the prediction range. That is, through various restrictions, exclude some unlikely items first, so that you can choose from a smaller number of target items. MSBR exists to meet this requirement. Through the information contained in many sessions, we will greatly reduce the prediction range. However, We don't have to treat these sessions equally. Difference between them is that their utilization value is not the same. Those sessions that are most relevant to the user's current session have greater utilization value. Therefore, in this procedure, we need a method to calculate the similarity between the candidate sessions and current session. This module will output the interests of the most similar TOP-{K} sessions, {K} is a hyperparameter, we will set different K values according to different datasets. 

Finally, based on the information provided in the above three modules, we will score all items in the Prediction Module, and select the items with the highest scores to recommend to users.

The overview of ENIREC model is shown in the figure 1.

\subsection{Pretrained GRU}
GRU plays an important role in our model, we have many small GRU modules responsible for capturing session information. If these GRU can't learn enough information, it will have a great negative impact on our model. Therefore, in order to stabilize the model, we must ensure that GRU can work. So we choose to use the sessions in the training set to pre-train a GRU first. GRU works as follows:
\begin{equation}
\begin{split}
    h_n &= GRU(i_{n},h_{n-1})\\
    i_n &\in s_t^{train}\\
    s_t^{train} &\in S_{train}\\
\end{split}
\end{equation}

The first hidden layer of GRU $h_0$ is initialized as a zero vector. Then take $h_0$ and $i_1$  as the first input of GRU, and input all items in $s_t^{train}$ in turn. We believe that the output of a layer in GRU represents the embedding of interests brought by those items in front of this layer. Therefore, the final output of GRU represents the interest of the whole session: 

\begin{equation}
    h^{s_t} = h_{\lvert s_t\rvert}
\end{equation}

\begin{figure}[ht]
 \centering
 \includegraphics[scale = 0.31]{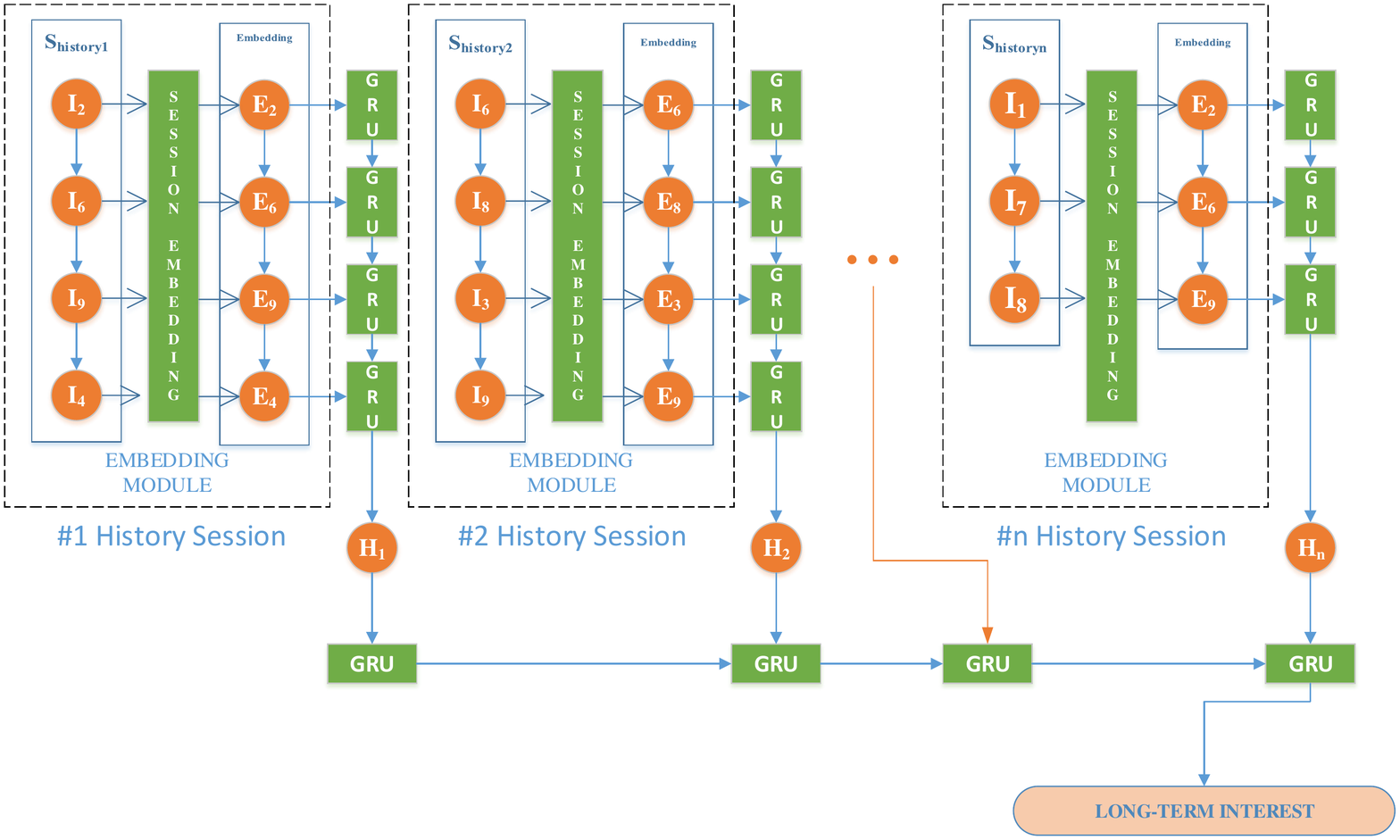}
 \caption{Way to extract long-term interest}
 \label{fig:LONG-TERM}
\end{figure}
\begin{figure*}[ht]
 \centering
 \includegraphics[height=0.55\textheight]{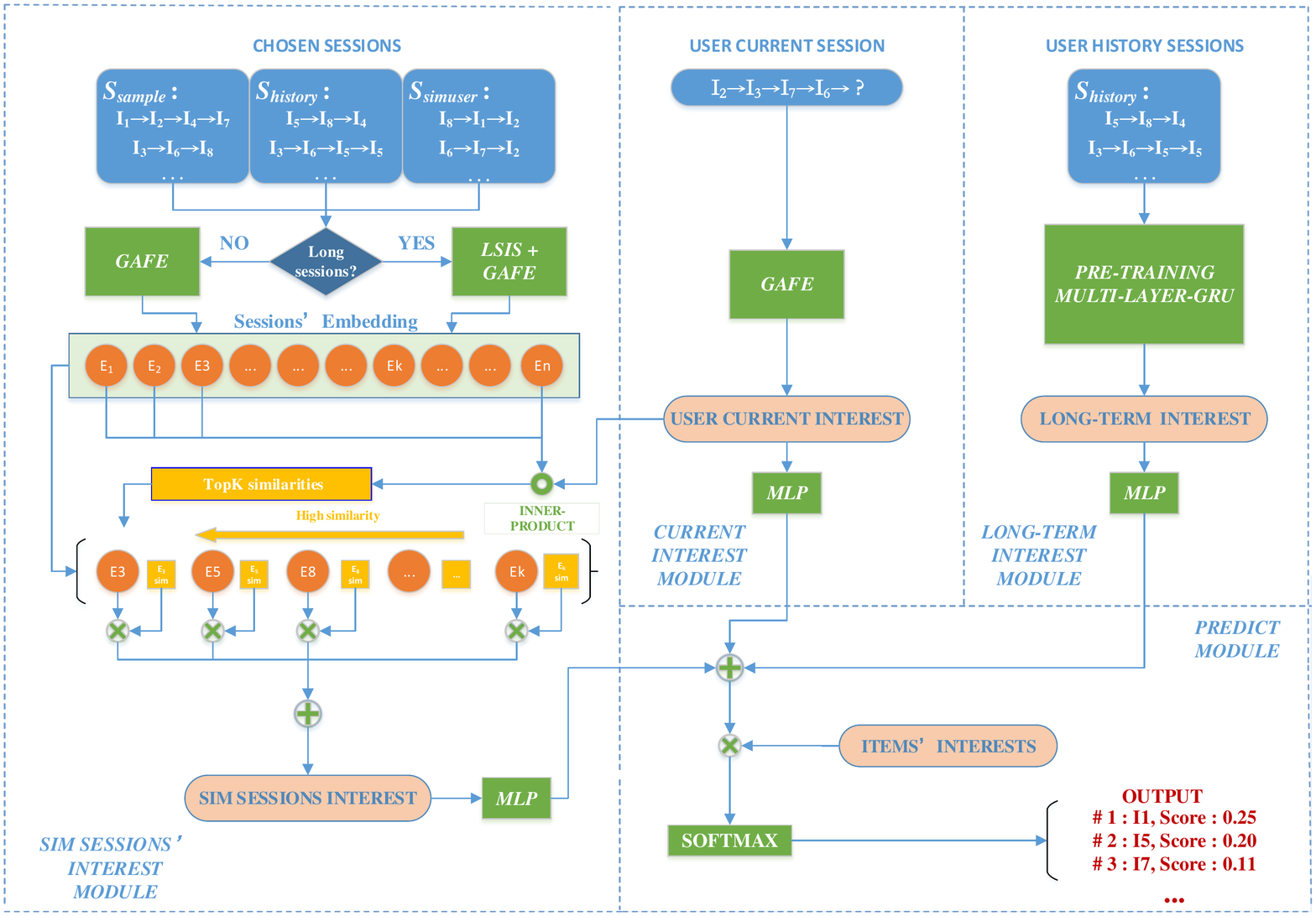}
 \caption{Overview of the proposed model ENIREC}
 \label{fig:ENIREC}
\end{figure*}
\subsection{Long-term Interest Module}
Learning long-term interest from the user's history has been proved to be one of the effective means\cite{28,31}. In SBR, there are many means to extract long-term interest, and RNN is commonly used. In this module, we build a multi-layer GRU. Any single-layer GRU can extract session interest. Multiple such GRU are spliced to extract the sequence relationship of sessions, that is, the user's long-term interest: 

\begin{equation}
\begin{split}
    E_{long-term-interest}&= GRU(h^{s_1^{his}},..., h^{s_{\lvert S_{history}\rvert}^{his}})\\
    s_t^{his}& \in S_{history}
\end{split}
\end{equation}

The reason why we don't apply GAFE instead of GRU in this module is that, unlike short sessions, people's long-term interests should pay more attention to a changing process, so the effect of GRU, the simple sequential processing method, should be better.

\subsection{Current Interest Module}
In this module, we will extract the users' current interests according to the users' current sessions. The current interests are the shortest-term interests we need to pay attention to and the most relevant information about the user's next item. Also, we look for similar sessions according to current interests. For the user's current session, we will directly use GAFE for processing:

\begin{equation}
\begin{split}
    E_{current-interest}&= GAFE(s_{current})
\end{split}
\end{equation}

The specific principle of GAFE will be described with formulas in detail in Section 4.4.1
\subsection{Sim Session's Interest Module}
Sim Session's Interest Module is the core module that distinguishes ENIREC from other models and makes the most contribution to improving the accuracy of recommendation. So we will introduce this module mainly with a large number of formulas and descriptions. 

Sim Session's Interest Module accepts and processes session data from three sets. This module includes three aspects.

\subsubsection{GRU-ATTENTION Feature Extractor}
\ 

\noindent First, Sim Session's Interest Module needs to have the basic ability to process short sessions. We introduce GRU-ATTENTION Feature Extractor(GAFE) to process short sessions. GAFE is a small module that we need to introduce it as a key point. It is precisely because of the excellent performance of GAFE ,which can extract session interest, that we can take the lead in recommendation. Different from other simple GRU, GAFE adopts the basic idea of attention, which desalinates the serialization relationship through the mechanism of attention. For a short session, the data extracted by attention will not be biased towards the last item. Instead, each item in the same session is balanced according to weight. So, it is more suitable for short sessions whose actual sequences are not important. 

To use attention, we first assign weights to each item in the sessions. The way to calculate the weight is to input the sessions into GRU, regard the input of each layer as the characteristic representation of the layer, calculate the similarity by inner product with the last layer's output, The formula is as follows:

\begin{equation}
\begin{split}
    H = \left\{ h_1,...,h_{\lvert s \rvert}\right\} &= GRU(s,h_0)\\
    w_n &= \langle h_n , h_{\lvert s \rvert}\rangle\\
    W &=\left\{w_1,...,w_{\lvert s \rvert}\right\}\\
\end{split}
\end{equation}

After that, We need to normalize the weight by softmax to get the weight, then just multiply it directly with the output of each layer and finally sum it to get the characteristic representation of the session:

\begin{equation}
\begin{split}
    W &= Softmax(W)\\
    E_s &= \sum_{n=1}^{\lvert s \rvert} w_n \times h_n\\
    w_n &\in W\\
    h_n &\in H
\end{split}
\end{equation}

\begin{figure*}
 \centering
 \subfloat[GAFE]
 {
  \centering          
  \includegraphics[scale=0.4]{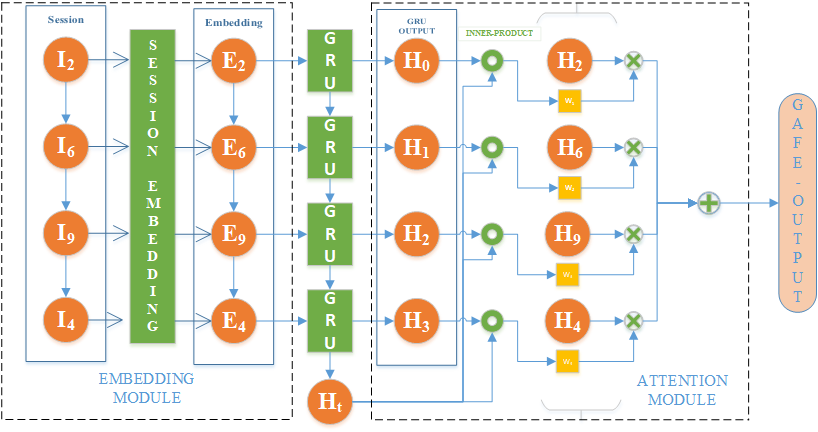}
 } 
 \subfloat[LSIS]
 {
  \centering          
  \includegraphics[scale=0.5]{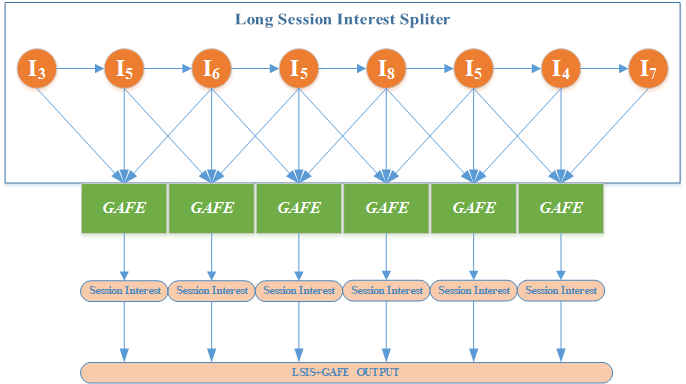}
 }
 \caption{Overview of GAFE and LSIS+GAFE}
 \label{fig:LSIS+GAFE}
\end{figure*}

\subsubsection{Long Session Interest Spliter}
\ 

\noindent Second, for long sessions, we use the method of LSIS to split the interests of long sessions into interests of multiple short sessions. The shorter the session, the more targeted the interest. For each long session

\begin{equation}
    s_{long}=\left\{i_1^{long},...,i_{\lvert s \rvert}^{long}\right\}
\end{equation}

The long session will be divided by sliding windows with a window size of 3, sliding from beginning to end on each long session, and intercepting the items in each window as a session:

\begin{equation}
\begin{split}
    s_{short_1}&= \left\{i_1^{long},i_2^{long},i_3^{long}\right\}\\
    &...\\
    s_{short_{{\lvert s \rvert}-2}}&=\left\{i_{{\lvert s \rvert}-2}^{long},i_{{\lvert s \rvert}-1}^{long},i_{\lvert s \rvert}^{long}\right\}\\
\end{split}
\end{equation}

In this way, a long session is divided into multiple short sessions with related relationships, which means that we disassemble the wide and multifaceted long-term interests that may be contained in a long session and turn them into many targeted short-term interests. Next, we can also use GAFE to process these short sessions:

\begin{equation}
    E_{long} = \left\{E_{short_1},...,E_{short_{{\lvert s \rvert}-2}}\right\}
\end{equation}

\subsubsection{Similar Sessions Processer}
\

\noindent Finally, We need to calculate the similarity between the alternative sessions and the user's current session, and take this similarity as the weight multiplied by the session, so as to distinguish the impact of different sessions on recommendation. The more similar the session, the greater the impact it will bring.

We need to use all input sessions, including short sessions splited by long sessions. After obtaining the embedding obtained after these sessions are processed by GAFE. We integrate these sessions' interest into a same large set $E_{sim}$

We will inner-product the embedding of these candidate sessions in $E_{sim}$ with the embedding of the users' current interests, so as to judge the degree of similarity. The results of inner-product will greater, if the degree of similarity of sessions is higher:

\begin{equation}
\begin{split}
    sim_n &= \langle e_n^{sim}, E_{current-interest} \rangle\\
    e_n^{sim} &\in E_{sim}\\
    sim_n &\in Sim
\end{split}
\end{equation}

Then we will multiply the embedding of these candidate sessions by their similarities, which can make the embedding value of more similar sessions larger, in another word, will increase the impact of the session for prediction. On the contrary, the embedding value of dissimilar sessions will be very small. Through the constraint of similarities, we can increase the positive impact and weaken the negative impact:

\begin{equation}
    e_n^{sim} = e_n^{sim} \times sim_n
\end{equation}

We do not consider using all candidate sessions, because the set of similar sessions is a very large set in real world datasets. Even if the impact of these sessions is reduced by multiplying the similarity, a large number of irrelevant sessions will still bring great noise to our prediction. In order to reduce the noise caused by sessions that we don't want to turn off, we don't consider using all sessions. We only select the sessions with top-k greater similarities $E_{sim}^* = \left\{ e_{1*}^{sim},...,e_{k*}^{sim}\right\}$('*' represents chosen). Finally, we add these selected sessions' embeddings as the output of this module.

\begin{equation}
\begin{split}
    E_{sim-session} = \sum_{n=1}^k e_{n*}^{sim}
\end{split}
\end{equation}

\subsection{Predict Module}

Firstly, each interest will be put in different Multi-Layer Perceptron(MLP) with only one hidden layer. Then we add the outputs of the three MLP as the embedding of next-item: 

\begin{equation}
\begin{split}
    E_{next-item} &= MLP_{cur}(E_{current-interest}) \\&+ MLP_{sim}(E_{sim-interest}) \\&+ MLP_{his}(E_{long-term-interest})
\end{split}
\end{equation}
Then we inner-product all items' embedding and next-item interest's embedding to give out the score of each item. $Score_{items}$. is a tensor with dimension $ 1 \times \lvert I \rvert $. Each item corresponds to a score in $Score_{items}$:
\begin{equation}
    Score_{items} = \langle E_{next-item}, E_{all-items}\rangle 
\end{equation}
At the end, the probability that user selects an item can be expressed as:
\begin{equation}
\begin{split}
    P(i | u_j) &= Softmax(Score_{items}^j)\\
\end{split}
\end{equation}

\section{EXPERIMENTS}
\subsection{Data Preparation}
In the experiment, we take two real-world datasets $Delicious$\cite{17} and $Reddit$\cite{12}, which are commonly used in SBRs. It is persuasive to achieve good results on these two datasets. $Delicious$ is a bookmark website. The dataset of it is a label dataset, which is often used for label recommendation. Reddit is a common dataset containing the interaction between users and items. 

In order to verify the effectiveness of our model, we will adopt the same processing method consistent with the existing work to process the two datasets. First, we will read all interactions of each user into a long sequence. Next, we will set a time threshold of 3600 seconds in our event, that is, the interaction occurring within 3600 seconds from the first interaction is determined to belong to one session, which turns the user's interaction into many sessions containing the user's adjacent interactions.
Then we delete too long and too short sessions. In the experiment, we delete sessions with length greater than 20 and length less than 2. After that, we calculate the occurrence frequency of each item and delete the items whose occurrence frequency are less than 10. And do the above length processing again for the sessions without these items. 

Then we divide the sessions of each user into training set, validation set and test set according to the proportion of 0-80\%, 70-80\% and 80-100\%, so that the test set, training set and verification set can have the user's information. In the following experimental details, we divide them by default according to the above proportion. We also divide according to the proportion of 0-70\%, 60-70\% and 70-100\% and 0-90\%, 80-90\% and 90-100\%, and the results are still better than the baseline methods, which will not be repeated here. The dataset parameters after processing are shown in the following table 1:

\begin{table}
  \caption{Dataset Parameters}
  \label{tab:freq}
\resizebox{\linewidth}{!}{
  \begin{tabular}{ccl}
    \toprule
     &Delicious&Reddit\\
    \midrule
    user\_num & 1,643 & 18,173\\
    item\_num & 5,005 & 13,521\\
    session\_num & 45,603 & 1,119,225\\
    session\_average\_length & 5.6 & 2.6\\
    session\_num per user & 27.8 & 61.6\\
  \bottomrule
\end{tabular}}
\end{table}

\subsection{Experiment Settings}
In order to intuitively compare the difference between our model and other existing models in the final experimental results. We will use the same evaluation method to compare our model. So we choose the widely used ranking metrics, i.e., Recall@20 and Recall@5, and Mean Reciprocal Rank MRR@20 and MRR@5 to evaluate the recommendation performance in the experiments.

\subsubsection{Baseline}
\

\noindent In order to compare the performance differences between our model and other existing models in many aspects, we carefully select models in three types for comparison: (1) single-session based (GRU4Rec, STAMP, SR-GNN), (2) multi-session based SBRSs (SKNN, HRNN, II-RNN) and (3) traditional sequential recommender systems SASRec and BERT4Rec).These are our baseline methods, some of them are very representative (covering Recurrent Neural Network, ATTENTION, Graph Neural Network and Memory Neural Network), and some of chosen models are state-of-the-art methods. 

\begin{itemize}

\item {\verb| GRU4Rec |}: A Recurrent Neural Network with Gated Recurrent Unit(GRU) is built for recommendations. GRU only gives recommendations based on a user's current session and the items' order in current session.\cite{12}

\item {\verb| STAMP |}: Considering the short-term interests of users, the Memory Neural Network is used to calculate the drift of a user's interest in real-time by using the ATTENTION mechanism according to capture a user's current clicks.\cite{11}

\item {\verb| SR-GNN |}: This is a single session based state-of-the art method proposed by the Chinese Academy of Sciences. This method shows the relationship between items in the session in the form of Graph Neural Network, and transmits information in the form of Gated GNN, so as to extract the user's preferences only from his current session.\cite{9}

\item {\verb| SKNN |}: It is an multi-session based method. By using the method of K-Nearest Neighbor. It makes recommendations by using the not only the user's current interest ,but also the information of similar sessions found from the whole dataset.\cite{26}

\item {\verb| HRNN |}: By constructing a heterogeneous Recurrent Neural Network, HRNN extracts interests from users' historical sessions and short-term interests from users' current sessions, and finally combines the two to make recommendations.\cite{16}

\item {\verb| II-RNN |}: II-RNN regards the user's last session and the current session slightly earlier as context relations, extracts their interests respectively, and regards them as a serialized preference change, so as to make recommendations.\cite{12}

\item {\verb| SASRec |}: SASRec is a typical model, which is a recommendation model based on Self-ATTENTION mechanism. It not only pays attention to user sessions, but also connects these sessions as user interest changes, and then predicts the user's next -- item according to the relationship between sessions.\cite{4}

\item {\verb| BERT4Rec|}: BERT4Rec is a state-of-the-art model, which directly migrates the recently popular BERT model in Natural Language Processing to Recommend System, and has achieved amazing results. The way to deal with the relationship between sessions is actually the same as SASRec.\cite{18}

\item {\verb| INSERT |}: INSERT is a state-of-the-art model. Its main sharing is to make use of more similar sessions, including user history sessions and similar user history sessions for recommendation. And by calculating the similarity for each session, the more similar the session, the greater the positive impact.\cite{24}

\end{itemize}
\begin{table*}[]
\caption{Results of Experiments}
\resizebox{\linewidth}{!}
{
\begin{tabular}{c|cccc|cccc}
\multicolumn{1}{l|}{} & \multicolumn{4}{c|}{\textbf{Delicious}}                                    & \multicolumn{4}{c}{\textbf{Reddit}}                                        \\ \hline
\multicolumn{1}{l|}{} & \textbf{Recall@5} & \textbf{Recall@20} & \textbf{MRR@5}  & \textbf{MRR@20} & \textbf{Recall@5} & \textbf{Recall@20} & \textbf{MRR@5}  & \textbf{MRR@20} \\ \hline
RNN                   & 0.1418            & 0.2716             & 0.0830          & 0.0957          & 0.1984            & 0.3544             & 0.1305          & 0.1458          \\
STAMP                 & 0.1476            & 0.2861             & 0.0861          & 0.0997          & 0.1534            & 0.2555             & 0.0981          & 0.1083          \\
SR-GNN                & 0.1680            & 0.3215             & 0.0931          & 0.1082          & 0.2377            & 0.4016             & 0.1555          & 0.1718          \\ \hline
SKNN                  & 0.1707            & 0.3487             & 0.0780          & 0.0960          & 0.1962            & 0.3758             & 0.0731          & 0.0912          \\
HRNN                  & 0.1749            & 0.3279             & 0.1038          & 0.1189          & 0.3482            & 0.5185             & 0.2436          & 0.2607          \\
II-RNN                & 0.1846            & 0.3493             & 0.1118          & 0.1279          & 0.3654            & 0.5481             & 0.2533          & 0.2717          \\ \hline
SASRec                & 0.1792            & 0.3431             & 0.0947          & 0.1104          & 0.3219            & 0.5711             & 0.1761          & 0.2012          \\
BERT4Rec              & 0.1755            & 0.3143             & 0.1096          & 0.1233          & \underline {0.4092}      & \textbf{0.6231}    & 0.2290          & 0.2518          \\
INSERT                & \underline {0.2163}      & \underline {0.3840}       & \underline {0.1278}    & \underline {0.1443}    & 0.3879            & 0.5588             & \underline {0.2684}    & \underline {0.2858}    \\ \hline
ENIREC                & \textbf{0.2516}   & \textbf{0.4159}    & \textbf{0.1558} & \textbf{0.1703} & \textbf{0.4410}   & \underline {0.5888}       & \textbf{0.3159} & \textbf{0.3421}
\end{tabular}
}

\end{table*}
\subsubsection{Parameters}
\

\noindent In terms of parameter settings, on the same parameters, we refer to the parameters in the above baseline methods and adjust our parameters to the same, so as to make a fair comparison between our methods.

\subsection{Evaluation and Analysis}
In this section, we will introduce the actual operation effect of ENIREC model, compare it with other baseline methods, and then analyze the advantages of ENIREC model according to the results of experiments.
\subsubsection{Performance of ENIREC}
\

\noindent We use the test results of these baseline models in Next item Recommendation in Short Sessions received in RECSYS 2021, and we run the INSERT model proposed in this paper, which is also the model with the best comprehensive effect , to prove that there is no problems in our parameters. Finally, our experimental results are shown in the figure below. For our ENIREC model, we run it many times and take the average of the running results of the model. We show the results in the table below:

\subsubsection{ENIREC vs Baseline methods}
\

\noindent The results show that ENIREC is better than the insert model with the best comprehensive effect in all aspects, which is enough to prove that the construction method of ENIREC model can effectively improve the recommendation accuracy.

First of all, compared with the traditional single-session based models GRU4Rec, STAMP and SR-GNN, it is obvious that multi-session based models can significantly improve the accuracy, which is also in line with the assumptions we mentioned earlier: more information, more accuracy. For GRU4Rec, the reasons for its poor performance also include that we mentioned earlier that RNN only focuses on the sequence relationship of sessions, but for short sessions, this type of relationship is not important in many cases. We can see from the comparison of the results of stamp and GRU4rec that the attention mechanism does improve the performance of the recommendation system. Based on this principle, we replaced the traditional GRU with GAFE. Similarly, SR-GNN is a state-of-the art model, which also breaks the sequence relationship and uses complex graph relationship to extract session information, which has been significantly improved. 

Our main comparison object is multi-session based models. It is not difficult for us to find a rule in these models. The more sessions are used, the higher the accuracy of recommendation will be. This also directly proves that multi-session based recommendation is a correct development direction. For SKNN, it search similar sessions from the whole dataset, but in fact, the similar sessions found by this method are not many and not stable, because its aim set is the whole dataset. Although the k-nearest neighbor method is adopted, it is still not good. Compared with SKNN, HRNN search from a very targeted set. It is the user's history session set, and the sessions in this set are probably related to user's current session. Therefore, it is more likely to find information related to this recommendation, and better results have been achieved in the results. II-RNN also improves efficiency by using the closest current session. And we can find that the effect of II-RNN is actually better than that of HRNN. From the analysis of the results, we believe that II-RNN focuses on short-term interest and HRNN focuses on long-term interest, which should be the main reason for the difference in results. This also confirms that there is nothing wrong with our focus on short-term interests.

Next, we will compare three special models, in which INSERT is the benchmark with our model. We will carefully compare the similarities and differences between the ENIREC and the three models, as well as the parts that produce differences in results. First, the task completed by SASRec is very similar to the function of Long-Term Internet Module in our model. They are not only concerned about sessions' interests, but also concatenate these interests as users' long-term interests. The difference is that SASRec adopts the Self-ATTENTION method. Next comes BERT4Rec, which has achieved excellent results in various fields. It can be predicted that it can be successful in the recommendation system. It is a general and effective method. 

The last model is INSERT, which is also the most important comparison model in this paper. The improvement made by INSERT compared with previous models is divided into two parts, which is also the reason why the comprehensive effect of INSERT is better than other models. The first point is that INSERT finds more sessions that can be used. Compared with the sessions found in other models, these sessions are not only more in number, but also more relevant. This is because INSERT not only looks for similar sessions from users' historical sessions, but also finds that users' similar users also contain many available sessions. The second improvement made by INSERT is that it does not blindly regard all sessions as sessions that can be used directly. It has a process of finding some really similar sessions. This process is to calculate the similarity between each session and the user's current session. Using the advantages of these two points, INSERT achieves very good results, and proves that these ideas are no problem. 

There are three main aspects that can be improved in INSERT, which can be learned from the comparison between the above baseline methods:
\begin{itemize}
\item We learned from SASRec that users' long-term interest is also a very important reference data, but INSERT did not consider it. Although INSERT take out users' history sessions separately as similar sessions, it did not capture the changes of users' history interests. For example, if the user bought high school books in the past and now college books, it means that the user has risen from high school to college. In INSERT, these are only two sessions that have no sequential relationship, and the insert model will not know this change. 
\item Insert obtains the useful information from the history sessions of users and similar users .And by calculating the similarity, dilute the influence of useless information. However, INSERT chose to consider these useless sessions as well. Limited to similarity, a single useless session will have little impact, but if all these sessions are considered, it will also bring noise that will greatly affect the recommendation results. And these are all historical information. Once the user's interest has a large offset, if it is fixed and only affected by these sessions, it will have an impact on the current session, which is completely different.  
\item For example, in the $Delicious$ dataset, INSERT focuses on solving the problem of a large number of short sessions, but ignores that there are 35.97\% of long sessions in $Delicious$. It is absolutely inappropriate to adopt the same processing method as short sessions for these long sessions. And RNN is still the processing method for short sessions in INSERT. It has been proved in the comparison between RNN, STAMP and SR-GNN that RNN does have disadvantages in the processing of short sequences.
\end{itemize}
For the above three points, ENIREC has also made three improvements:
\begin{itemize}
\item In the Long-Term module, we have established a multi-layer GRU to stably extract users' long-term interests ,to make sure that we can make good use of the information of long-term interest.
\item In the process of searching for similar sessions, we added the selection of sampling in the whole dataset. The selection of this part can add a disturbance to the selection of sessions with all historical data, and cooperate with the way that we only take the most similar sessions in TOPK. We can only use the most similar sessions and avoid the noise caused by different sessions.
\item For long sessions in the datasets, we propose the LSIS method to ensure that the information of long sessions can be better used, and for short sessions, we adopt the GAFE , which can help us better extract the interest of short sessions.
\end{itemize}

\subsection{Ablation Analysis}
For the above three changes, we propose three different versions of ENIREC to prove the effectiveness of the above three changes.For each case, we conducted several experiments and took the average value.

\subsubsection{ENIREC-a}
\

\noindent For the first improvement of benchmarking, ENIREC-a can directly judge whether the Long-Term Interest Module has a positive impact on the recommendation results by removing the long term interest module, and can judge whether the module is indispensable to ENIREC.

We choose to conduct comparative experiments on $Delicious$, and the results are shown in the table below. We can clearly draw the conclusion that Long-Term Interest Module has effectively improved the accuracy of recommendation.

\subsubsection{ENIREC-b}
\

\noindent For the second improvement of benchmarking, We choose to conduct comparative experiments on $Delicious$, and the results are shown in the table below. We tried two ways, one is ENIREC-b-1. This method is to directly remove the operation of selecting TOPK, that is, we will directly use all alternative sessions. 

The second is ENIREC-b-2 to retain the operation of selecting TOPK, but cancel the method of sampling sessions from the dataset.

From the table, we can infer that adding sample sessions will add disturbance. If we do not control this disturbance by taking TOPK, it will affect the recommendation accuracy. If we only select TOPK, we will find that the accuracy has increased, but the increase is not obvious, and it has decreased in the other two evaluation standards, so we can draw a conclusion, Combining these two methods can achieve good results.

\subsubsection{ENIREC-c}
\
The third is that we directly replace all GAFE in the model with GRU. The results are shown in the table below. Eventually, We find that GAFE can significantly improve the performance compared with GRU.

This also proves that directly using RNN to process session-based recommendation data is not a perfect way.

\begin{table}[]
\caption{Ablation Results}
\resizebox{\linewidth}{!}{
\begin{tabular}{c|cccc}
           & \textbf{RECALL@5} & \textbf{RECALL@20} & \textbf{MRR@5}  & \textbf{MRR@20} \\ \hline
ENIREC-a   & 0.2274            & 0.3829             & 0.1375          & 0.1530          \\
ENIREC-b-1 & 0.1970            & 0.3531             & 0.1169          & 0.1411          \\
ENIREC-b-2 & 0.2210            & 0.4015             & 0.1451          & 0.1610          \\
ENIREC-c   & 0.2191            & 0.3783             & 0.1333          & 0.1489          \\ \hline
ENIREC     & \textbf{0.2416}   & \textbf{0.4159}    & \textbf{0.1558} & \textbf{0.1703}
\end{tabular}}
\end{table}

\section{Conclusion}
ENIREC emphasizes extracting users' long-term interests and current interests, and comprehensively considers helping recommendation. ENIREC also finds sessions similar to users' current sessions from various sessions, and uses the information of these sessions to help recommendation by calculating similarity for every session and taking TOPK of them.
At the same time, the huge noise caused by these sessions is avoided. 

In a word, the experimental results show that facing the limitations of short session, ENIREC can still successfully make use of the advantages of short sessions and avoid the disadvantages of short session, so as to achieve satisfactory results in the final recommendation accuracy. 

Also, through the comparison between ablation experiment and other baseline experiments, we found that the starting point of ENIREC improvement is very correct. It is proved that each module of ENIREC plays a positive role in improving the accuracy of recommendation.




\end{document}